\documentclass[10pt,a4paper]{article}
\usepackage[T2A]{fontenc}
\usepackage[utf8]{inputenc}
\usepackage[english]{babel}
\usepackage{amssymb,graphicx}
\usepackage{amsmath,amsfonts}
\usepackage{amsthm}
\usepackage{framed}
\usepackage{fullpage}
\usepackage[makeroom]{cancel}
\usepackage[export]{adjustbox} % Adjust the vertical alignment of the
\usepackage{microtype}
\usepackage{hyperref}

\title{\textbf{$\mathfrak{gl}_N$ Higgsed networks}}
\author{Yegor Zenkevich\thanks{yegor.zenkevich@gmail.com}\\
  {\small\textit{SISSA, via
      Bonomea 265, 34136 Trieste, Italy,}}\\
  {\small\textit{INFN, Sezione di Trieste,}}\\
  {\small\textit{IGAP, via Beirut 2/1, 34151 Trieste, Italy,}}\\
  {\small\textit{ITEP, Bolshaya Cheremushkinskaya street 25, 117218
      Moscow, Russia,}}\\
  {\small\textit{ITMP MSU, Leninskie gory 1, 119991 Moscow, Russia,}}\\
  {\small\textit{MIPT, Institutskii pereulok 9, 141700, Dolgoprudny, Russia,}}}
\date{}
\begin{document}
\maketitle
\vspace{-45ex}
\begin{flushright}
  ITEP-TH-38/19
\end{flushright}
\vspace{35ex}

\begin{abstract}
  We generalize the framework of Higgsed networks of intertwiners to
  the quantum toroidal algebra associated to Lie algebra
  $\mathfrak{gl}_N$. Using our formalism we obtain a systems of
  screening operators corresponding to $W$-algebras associated to
  toric strip geometries and reproduce partition functions of $3d$
  theories on orbifolded backgrounds.
\end{abstract}

\section{Introduction}
\label{sec:introduction}
Quantum toroidal algebras~\cite{Ding-Iohara},~\cite{Miki-07} are
quantum deformations of double loop algebras. They are natural from
several points of view and have intricate representation theory. Most
importantly for us, they play a crucial role in different versions of
the AGT correspondence~\cite{Alday:2009aq}, which in its abstract
form~\cite{Nakajima} is a statement about actions of certain algebras
on the spaces of BPS objects in gauge theories. In many cases the
algebras featuring in the correspondence are quantum toroidal algebras
and the spaces of BPS states are $K$-theories of various moduli
spaces. Here we will not be concerned with the mechanism of the action
itself, which belongs to the geometric representation theory (see
e.g.~\cite{Carlsson:2013jka}). We will use the representation theory
of the algebras to build certain intertwining operators and screening
currents and with their help compute partition functions of $3d$ gauge
theories. In doing so we follow closely our previous
work~\cite{Z-higgsed}, where we have considered similar problem for
the algebra $U_{q_1,q_2,q_3}(\widehat{\widehat{\mathfrak{gl}}}_1)$.

The intertwiners we are going to construct are between horizontal Fock
representations and vertical vector representations of the quantum
toroidal algebra
$U_{q_1,q_2,q_3}(\widehat{\widehat{\mathfrak{gl}}}_N)$. We will see
that, just as in the ``abelian''
$U_{q_1,q_2,q_3}(\widehat{\widehat{\mathfrak{gl}}}_1)$ case, these
intertwiners are convenient building blocks which can be glued
together into what we have called a Higgsed network. The idea of the
construction is as follows. In~\cite{AFS} the intertwining operators
between Fock representations of
$U_{q_1,q_2,q_3}(\widehat{\widehat{\mathfrak{gl}}}_1)$ have been shown
to reproduce refined topological vertex computations, which in turn
can be used to compute Nekrasov partition
functions~\cite{Nekrasov:2002qd} of $5d$ gauge theories on
$\mathbb{C}^2 \times S^1$. In these $5d$
gauge theories one can tune the parameters so that the theory becomes
dual to a $3d$ theory living on the worldvolume of vortices in the
Higgs phase. This tuning procedure is known as Higgsing. The Higgsed
intertwiners provide an algebraic framework, in which correlators
automatically reproduce Higgsed theory partition functions. Since a
lot of degrees of freedom decouple during the Higgsing, the
representations featuring in the Higgsed intertwiners are ``smaller''
than the Fock representations from~\cite{AFS}. In these smaller vector
representations the states are labelled by integers (cf.\ with the
Fock representations where they are labelled by Young diagrams, or
tuples of integers). Since the representations are smaller, the
intertwiners are easier to handle and the computations become more
transparent. Besides, vector representations allow one to better use
the full power of the large automorphism group of the quantum toroidal
algebra (for details see~\cite{Z-higgsed}).

In the present short note we consider the following variation of the
Higgsing procedure. Let the original $5d$ gauge theory live not on
$\mathbb{C}^2\times S^1$, but on an ALE space $\mathbb{C}^2/
\mathbb{Z}_N\times S^1$. In the algebraic language the partition
function of the $5d$ theory corresponds to a network of Fock space
intertwiners for the algebra
$U_{q_1,q_2,q_3}(\widehat{\widehat{\mathfrak{gl}}}_N)$~\cite{Awata:2017lqa}\footnote{More
  general algebraic models corresponding to the $5d$ theory on
  $\mathbb{C}^2/(\mathbb{Z}_{N_1}\times \mathbb{Z}_{N_2}) \times S^1$
  have recently been considered~\cite{Bourgine:2019phm}.}. The tuning
of the parameters corresponding to Higgsing can still be performed,
and one arrives at a $3d$ theory on the vortex worldvolume. However,
now the theory is ``orbifolded'': it lives on
$\mathbb{C}/\mathbb{Z}_N\times S^1$ and there is an additional
$\mathbb{Z}_N$ $R$-symmetry twist when going around the orbifold
point. We will obtain the partition function of the orbifolded theory
from the network of Higgsed
$U_{q_1,q_2,q_3}(\widehat{\widehat{\mathfrak{gl}}}_N)$
intertwiners. It turns out that the form of the partition function and
the its algebraic properties are quite similar to those of the
original theory. The main difference is that the partition function is
not just a function, but a vector-valued function, i.e.\ there are
several partition functions labelled by the indices associated with
fundamental hypermultiplets (external lines in the intertwiner
picture).

The plan of the paper is as follows. We construct the intertwiners in
sec.~\ref{sec:intertwiners}, compute their correlators in
sec.~\ref{sec:correlators} and show how to interpret them in the gauge
theory and algebraic language in
sec.~\ref{sec:screenings-networks}. Since, the main ideas have already
been laid out in~\cite{Z-higgsed}, we allow ourselves to be brief and
present only the main features of the $\mathfrak{gl}_N$
generalization.

In order not to clutter the main part of the paper with technical
details, we move the notations and definitions to the
appendices. There we define the algebra
$U_{q_1,q_2,q_3}(\widehat{\widehat{\mathfrak{gl}}}_N)$
(sec.~\ref{sec:algebra}), list its properties
(sec.~\ref{sec:coproduct-structure}, \ref{sec:autom-algebra} and
\ref{sec:algebra-as-quantum}), and write down its relevant
representations (sec.~\ref{sec:vect-repr-u},
\ref{sec:fock-representation-}). Sec.~\ref{sec:branching-into-}
contains some facts about the correspondence between the
$U_{q_1,q_2,q_3}(\widehat{\widehat{\mathfrak{gl}}}_N)$ algebra and
geometry of a class of toric CY threefolds.

\section{The intertwiners}
\label{sec:intertwiners}
Similarly to~\cite{Z-higgsed} we would like to build an intertwining
operator $\Phi: \mathcal{V}_{q_1} \otimes
\mathcal{F}^{(1,0),k}_{q_1,q_3}(u) \to
\mathcal{F}^{(1,0),k}_{q_1,q_3}(q_3^{-1} u)$. By definition the
intertwiner satisfies
\begin{equation}
  \label{eq:58}
  g \Phi = \Phi \Delta(g),
\end{equation}
where $g$ is any element of
$U_{q_1,q_2,q_3}(\widehat{\widehat{\mathfrak{gl}}}_N)$ and $\Delta$ is
the coproduct given in Appendix~\ref{sec:coproduct-structure}.  Representations $\mathcal{V}_{q_1}$ and
$\mathcal{F}^{(1,0),k}_{q_1,q_3}(u)$ featuring in the intertwiner are
defined in Appendices~\ref{sec:vect-repr-u}
and~\ref{sec:fock-representation-} respectively. Since there will be
only one Fock representation in our story, we omit the indices
$_{q_1,q_3}$ henceforth.

We represent the intertwiner graphically as a trivalent junction:
\begin{equation}
  \label{eq:32}
  \Phi^j_{q_1} (w) =\quad  \includegraphics[valign=c]{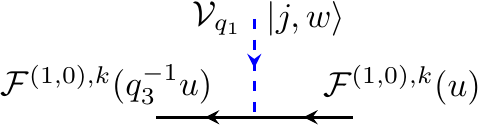}, \qquad j = 1,\ldots, N.
\end{equation}
We denote the vector representation $V_{q_1}$ as blue dashed line,
while the Fock space is represented by a solid black line. The
essential new ingredients compared to the
$U_{q_1,q_2,q_3}(\widehat{\widehat{\mathfrak{gl}}}_1)$ case are the
$\mathbb{C}^N$ index $j$ of the intertwiner and the color $k$ of the
horizontal Fock representation. The explicit expression for the
intertwiner is
\begin{multline}
  \label{eq:35}
  \Phi^j_{q_1} (w) =  e^{Q_j} (- q^{-1} w)^{P_j} \exp
  \left[ \sum_{n\geq 1} \frac{w^n}{n}
    q^{-n} a_{j, -n} \right] \times \\
  \times \exp \left[ - \sum_{n\geq 1} \frac{w^{-n}}{n} \left( q^n
      a_{j, n} + \frac{q^{-n} - q^n}{1 - q_1^{n N}}\sum_{l=0}^{N-1}
      q_1^{n (\overline{l-j})} a_{l,n} \right) \right],
\end{multline}
where $\overline{l-j} = (l-j) \bmod N$. The definitions of the bosons
$a_{j,n}$ and the zero modes $Q_j$, $P_j$ are given in
Appendix~\ref{sec:indep-heis-gener}. Notice that in the ``unrefined
limit'' $q \to 1$ the second line simplifies, so that $\Phi^j_{q_1}
(w)$ becomes the conventional undeformed free boson vertex operator.

The dual intertwiner $\Phi_j^{*,q_1, k} :
\mathcal{F}^{(1,0),k}_{q_1,q_3}(u) \to
\mathcal{F}^{(1,0),k}_{q_1,q_3}(q_3 u) \otimes \mathcal{V}_{q_1}$ is
given by
\begin{multline}
  \label{eq:56}
  \Phi_j^{*,q_1} (w)
  =\quad    \includegraphics[valign=c]{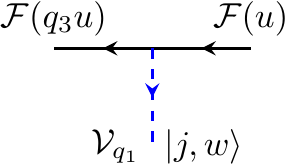}\quad
  =  e^{-Q_j} (- q^{-1} w)^{-P_j + \frac{1}{2} (1
    - \beta)} \exp \left[ - \sum_{n\geq 1} \frac{w^n}{n}
    a_{j, -n} \right]\times \\
  \times \exp \left[ \sum_{n\geq 1} \frac{w^{-n}}{n} q^n \left( q^n
      a_{j, n} + \frac{q^{-n} - q^n}{1 - q_1^{n N}}\sum_{l=0}^{N-1}
      q_1^{n (\overline{l-j})} a_{l,n} \right) \right].
\end{multline}
where $\beta = \frac{\ln q_3}{\ln q_1}$.

Due to the reflection symmetry $\rho$ of the algebra (see
Appendix~\ref{sec:autom-algebra}) there is another pair of
intertwiners with the vector representation $\mathcal{V}_{q_3}$
(denoted by the dashed red lines):
\begin{multline}
  \label{eq:57}
  \Phi^j_{q_3} (w) =
  \quad \includegraphics[valign=c]{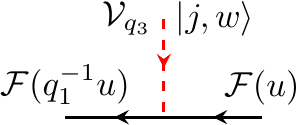} \quad =
  e^{-Q_{\overline{j+1}}} (- q^{-1} w)^{-P_{\overline{j+1}}} \times\\
  \times \exp \left[ - \sum_{n \geq 1} \frac{w^n}{n}
    d^{-n} q_2^{-n} \left( a_{\overline{j+1},-n} + q_3^n \frac{1 -
          q_2^n}{1 - q_3^{nN}} \sum_{l=0}^{N-1}
        q_3^{\overline{(l-j-2)} n} a_{l,-n} \right) \right] \exp
    \left[ \sum_{n \geq 1} \frac{w^{-n}}{n} d^n a_{\overline{j+1},n} \right],
\end{multline}
\begin{multline}
\label{eq:67}
  \Phi^{*,q_3}_j (w) =
  \quad \includegraphics[valign=c]{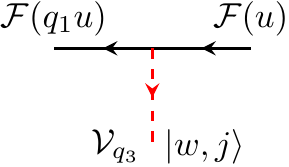}
  \quad = e^{Q_{\overline{j+1}}} (- q^{-1} w)^{P_{\overline{j+1}} +\frac{1}{2}(1 +
    \beta)} \times\\
  \times \exp \left[ \sum_{n \geq 1} \frac{w^n}{n} q_3^n \left(
      a_{\overline{j+1},-n} + q_3^n \frac{1 - q_2^n}{1 - q_3^{nN}}
          \sum_{l=0}^{N-1} q_3^{\overline{(l-j-2)} n} a_{l,-n} \right)
      \right] \exp \left[ - \sum_{n \geq 1} \frac{w^{-n}}{n} q_3^{-n}
        a_{\overline{j+1},n} \right].
\end{multline}
We will omit some zero modes henceforth since their contributions to
the correlators are just powers of the spectral parameters.

\section{Correlators}
\label{sec:correlators}
Having the explicit form of the intertwining
operators~\eqref{eq:35}--\eqref{eq:67} we can straightforwardly
compute their correlators. Moreover, since the intertwiners are built
free boson vertex operators Wick theorem applies, so it is sufficient
to compute only the pairwise correlators. They are given by
\begin{equation}
  \label{eq:45}
   \includegraphics[valign=c]{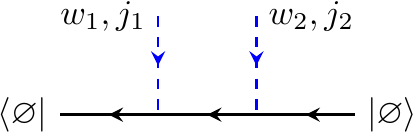} \quad = \langle \varnothing |  \Phi^{j_1}_{q_1,k}(w_1) \Phi^{j_2}_{q_1,k}(w_2) |\varnothing
  \rangle   \sim (w_1 - w_2)^{\delta_{j_1,j_2}} \frac{\left( q_1^{\overline{j_2 - j_1}} q_2^{-1} \frac{w_2}{w_1} ; q_1^N
    \right)_{\infty}}{\left( q_1^{\overline{j_2 - j_1}} \frac{w_2}{w_1} ; q_1^N \right)_{\infty}},
\end{equation}
\begin{equation}
  \label{eq:49}
   \includegraphics[valign=c]{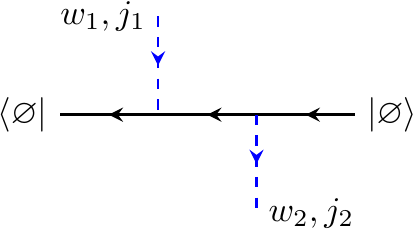} \quad   = \langle \varnothing |  \Phi^{j_1}_{q_1,k}(w_1) \Phi_{j_2}^{*,q_1,k}(w_2) |\varnothing
  \rangle   \sim (w_1 - q w_2)^{-\delta_{j_1,j_2}} \frac{\left( q_1^{\overline{j_2 - j_1}} q \frac{w_2}{w_1} ; q_1^N
    \right)_{\infty}}{\left( q_1^{\overline{j_2 - j_1}} q^{-1} \frac{w_2}{w_1} ; q_1^N \right)_{\infty}},
\end{equation}
\begin{equation}
  \label{eq:50}
   \includegraphics[valign=c]{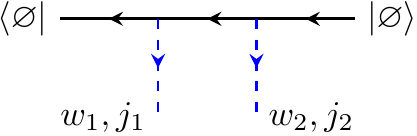} \quad = \langle \varnothing |  \Phi_{j_1}^{*,q_1,k}(w_1) \Phi_{j_2}^{*,q_1,k}(w_2) |\varnothing
  \rangle   \sim (w_1 - q_2 w_2)^{\delta_{j_1,j_2}} \frac{\left( q_1^{\overline{j_2 - j_1}} \frac{w_2}{w_1} ; q_1^N
    \right)_{\infty}}{\left( q_1^{\overline{j_2 - j_1}} q_2\frac{w_2}{w_1} ; q_1^N \right)_{\infty}},
\end{equation}
\begin{equation}
\label{eq:51}
   \includegraphics[valign=c]{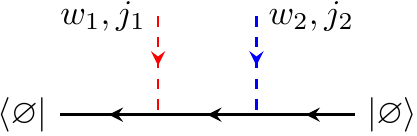}\quad = \langle \varnothing |  \Phi^{j_1}_{q_3,k}(w_1) \Phi^{j_2}_{q_1,k}(w_2) |\varnothing
\rangle   \sim (w_1 - q_1 w_2)^{-\bar{\delta}_{j_1+1,j_2}},
\end{equation}
\begin{equation}
  \label{eq:52}
    \includegraphics[valign=c]{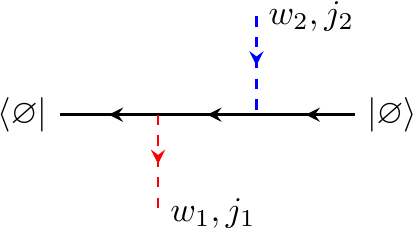}
   \quad = \langle \varnothing |  \Phi_{j_1}^{*,q_3,k}(w_1) \Phi^{j_2}_{q_1,k}(w_2) |\varnothing
\rangle   \sim (w_1 - d w_2)^{\bar{\delta}_{j_1+1,j_2}} ,
\end{equation}
\begin{equation}
  \label{eq:55}
  \includegraphics[valign=c]{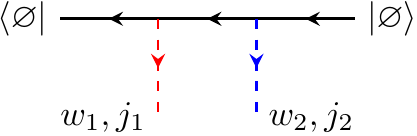}
  \quad = \langle \varnothing |  \Phi_{j_1}^{*,q_3,k}(w_1) \Phi_{j_2}^{*,q_1,k}(w_2) |\varnothing  \rangle   \sim (w_1 - q_3^{-1} w_2)^{-\bar{\delta}_{j_1+1,j_2}}.
\end{equation}
The proportionality signs in Eqs.~\eqref{eq:45}--\eqref{eq:55} are to
remind us that we have omitted some simple factors originating from
the zero modes. We don't list the correlators of $\mathcal{V}_{q_3}$
intertwiner with another $\mathcal{V}_{q_3}$ intertwiner, because they
are obtained from Eqs.~\eqref{eq:45}--\eqref{eq:50} by the exchange
$q_1 \leftrightarrow q_3$ and $j \leftrightarrow \overline{(-j)}$.
  
\subsection{$R$-matrix}
\label{sec:networks-r-matrices}
Having computed the correlators, we can deduce the commutation
relations for the intertwiners. They feature the diagonal $R$-matrix
of $U_{q_1,q_2,q_3}(\widehat{\widehat{\mathfrak{gl}}}_N)$ acting in
the tensor product of two vector representations:
\begin{equation}
  \label{eq:54}
  \Phi^{j_1}_{q_1,k}(w_1) \Phi^{j_2}_{q_1,k}(w_2) = A \left(
    \frac{w_1}{w_2} \right) R_{q_1 q_1}\left( \frac{w_1}{w_2} \right)
  \Phi^{j_2}_{q_1,k}(w_2) \Phi^{j_1}_{q_1,k}(w_1) ,
\end{equation}
where $A (x)$ is a certain $q_1^N$-periodic
combination of theta-functions and the $R$-matrix is given by
\begin{equation}
  \label{eq:59}
  R(x) = \left( \frac{w_2 - q_2 w_1}{w_2 - w_1} \right)^{\delta_{j_1,j_2}} \frac{\left( q_1^{\overline{j_1 - j_2}} \frac{w_1}{w_2}
      ;q_1^N \right)_{\infty}^2}{\left( q_1^{\overline{j_1 - j_2}} q_2 \frac{w_1}{w_2} ;q_1^N \right)_{\infty}\left( q_1^{\overline{j_1 - j_2}} q_2^{-1} \frac{w_1}{w_2} ;q_1^N \right)_{\infty}}.
\end{equation}
Similarly, we learn that
\begin{equation}
  \label{eq:60}
  \Phi_{j_1}^{*,q_1,k}(w_1) \Phi_{j_2}^{*,q_1,k}(w_2) = B \left(
    \frac{w_1}{w_2} \right) \frac{1}{R_{q_1 q_1}\left( \frac{w_1}{w_2} \right)}\Phi_{j_2}^{*,q_1,k}(w_2) \Phi_{j_1}^{*,q_1,k}(w_1)  ,
\end{equation}
where $B (x)$ is another $q_1^N$-periodic combination. The dual
intertwiners commute (up to a $q_1^N$-periodic function $C(x)$)
\begin{equation}
  \label{eq:61}
  \Phi_{j_1}^{*,q_1,k}(w_1) \Phi^{j_2}_{q_1,k}(w_2) = C \left(
    \frac{w_1}{w_2} \right)  \Phi^{j_2}_{q_1,k}(w_2) \Phi_{j_1}^{*,q_1,k}(w_1).
\end{equation}
We don't pay attention to the $q_1^N$-periodic functions
since in the partition function we are going to study in the next
section the intertwiners are going to be under the $q_1^N$-Jackson
integral (or, equivalently, under contour integral with
$q_1^N$-periodic strings of poles).

\section{Screenings and networks}
\label{sec:screenings-networks}

Similarly to~\cite{Z-higgsed} we can build a network of intertwiners
by combining them either vertically or horizontally. We have
considered all possible horizontal combinations in the previous
section. In this section we combine the intertwiners vertically to tie
together several horizontal solid lines.

\subsection{Screening operators}
\label{sec:screening-operators}
Two possible vertical combinations of intertwiners give rise to two
screening currents acting in the tensor product of two Fock spaces:
\begin{multline}
  \label{eq:62}
  S^j_{q_1}(w) = \quad\includegraphics[valign=c]{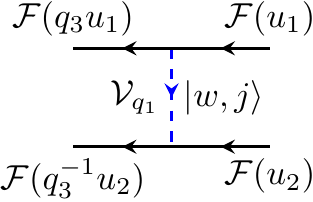}
  \quad \sim
  \exp \left[ \sum_{n\geq 1} \frac{w^n}{n}
    q^{-n}( a^{(2)}_{j, -n} - q^n a^{(1)}_{j, -n}) \right] \times \\
  \times \exp \left[ - \sum_{n\geq 1} \frac{w^{-n}}{n} \left( q^n
      (a^{(2)}_{j, n} - q^n a^{(1)}_{j, n}) + \frac{q^{-n} - q^n}{1 - q_1^{n
      N}}\sum_{l=0}^{N-1} q_1^{n (\overline{l-j})} (a^{(2)}_{l,n} - q^n a^{(1)}_{l, n})
\right) \right],
\end{multline}
\begin{multline}
  \label{eq:63}
  S^j_{q_3}(w) =
  \quad\includegraphics[valign=c]{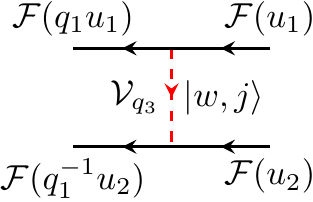}
  \quad \sim \\
  \sim \exp \left[ - \sum_{n \geq 1} \frac{w^n}{n} d^{-n} q_2^{-n}
    \left( (a^{(2)}_{\overline{j+1},-n} - q^n
          a^{(1)}_{\overline{j+1},-n}) + q_3^n \frac{1 - q_2^n}{1 -
              q_3^{nN}} \sum_{l=0}^{N-1} q_3^{\overline{(l-j-2)} n}
            (a^{(2)}_{l,-n} - q^n a^{(1)}_{\overline{j+1},-n}) \right)
            \right]
            \times \\
            \times \exp \left[ \sum_{n \geq 1} \frac{w^{-n}}{n} d^n
              (a^{(2)}_{\overline{j+1},n} - q^n
                  a^{(1)}_{\overline{j+1},n}) \right],
\end{multline}
where $a^{(1)}_{i,n}$ and $a^{(2)}_{i,n}$ act on the first and the
second Fock spaces respectively.  If we sum over a complete basis of
states in the intermediate vector representations in
Eqs.~\eqref{eq:62},~\eqref{eq:63} we will get intertwining operators
acting in the tensor product of two Fock spaces:
\begin{align}
  \label{eq:64}
  Q_{q_1} &= \sum_{j=0}^{N-1} \int \frac{d_{q_1^N}w}{w}\, S^j_{q_1}(w) =
  \sum_{j=0}^{N-1} \oint \frac{dw}{w}\, S^j_{q_1}(w),\\
  Q_{q_3} &= \sum_{j=0}^{N-1} \oint \frac{d_{q_1^N} w}{w}\,
  S^j_{q_3}(w)=\sum_{j=0}^{N-1} \oint \frac{dw}{w}\,
  S^j_{q_3}(w)\label{eq:65}
\end{align}
In other words, integrated combinations like $Q_{q_1}$ and $Q_{q_3}$
are screening charges commuting with the action of the whole
$U_{q_1,q_2,q_3}(\widehat{\widehat{\mathfrak{gl}}}_N)$ algebra. We can
use these screening operators to build ``$\mathfrak{gl}_N$-extended''
$W_k$ algebras acting on the tensor product of $k N$ Fock
spaces\footnote{A version of spectral duality can be used to exchange
  $k$ and $N$ as was demonstrated in~\cite{tor-spectral}.}. By Fock
spaces here we mean representations of a single boson, and denote them
by $\mathfrak{F}$ to distinguish from the Fock representation
$\mathcal{F}$ of
$U_{q_1,q_2,q_3}(\widehat{\widehat{\mathfrak{gl}}}_N)$ which is
generated by $N$ bosons. To build the $W_k$-algebra one needs to
cluster together $N$-tuples of Fock spaces:
\begin{equation}
  \label{eq:73}
  \mathfrak{F}^{\otimes k N}  \simeq (\mathfrak{F}^{\otimes
    N})^{\otimes k} \simeq \mathcal{F}^{\otimes k}.
\end{equation}
There are $k$ tuples, on which the screening charges $Q_{q_1}^{(a)}$
and $Q_{q_3}^{(a)}$ ($a=1,\ldots, k-1$) can act. The screenings
$Q_{q_1}^{(a)}$ and $Q_{q_3}^{(a)}$ act between $a$-th and $(a+1)$-th
$N$-tuple and the $W_k$-algebra is generated by the operators
commuting with all the screenings. There is a constructive way to
obtain the generators: one can simply take a coproduct of any element
of $U_{q_1,q_2,q_3}(\widehat{\widehat{\mathfrak{gl}}}_N)$ and evaluate
it on the tensor product of $k$ Fock spaces. It will automatically
commute with the screening charges. As an example of the
``$\mathfrak{gl}_N$-extended'' $q$-deformed $W_2$ generators we take
$\Delta(E_i(z))$ (see Appendices~\ref{sec:coproduct-structure}
and~\ref{sec:fock-representation-} for the definition of the coproduct
and the Fock representation respectively):
\begin{multline}
  \label{eq:74}
  \Delta(E_i(z))|_{\mathcal{F}^{k_1}(u_1)\otimes
    \mathcal{F}^{k_2}(u_2)} = e^{\delta_{i,k_1}P^{(1)}} e^{Q^{(1)}_i - Q^{(1)}_{\overline{i+1}}} z^{P^{(1)}_i -
        P^{(1)}_{\overline{i+1}}+1} (-d)^{P^{(1)}_{\overline{i+1}}} \exp \left[ \sum_{n \geq 1}
              \frac{z^n}{n}q^{-n} ( a^{(1)}_{i,-n} - q_1^{-n} a^{(1)}_{\overline{i+1},-n} )\right]\times \\
                \times \exp \left[ - \sum_{n \geq 1}
                  \frac{z^{-n}}{n}q^{-n} ( a^{(1)}_{i,n} - q_3^{-n}
                  a^{(1)}_{\overline{i+1},n} )\right] +
                    e^{\delta_{i,k_2}P^{(2)}} e^{Q^{(2)}_i -
                      Q^{(2)}_{\overline{i+1}}} (q z)^{P^{(2)}_i -
        P^{(2)}_{\overline{i+1}}+1} (-d)^{P^{(2)}_{\overline{i+1}}}
            q^{-P^{(1)}_i + P^{(1)}_{\overline{i+1}}} \times \\
                \times  \exp \left[ \sum_{n \geq 1}
              \frac{z^n}{n} \left( a^{(2)}_{i,-n} - q_1^{-n}
                a^{(2)}_{\overline{i+1},-n} -(q^n
                    - q^{-n})q^{\frac{n}{2}} \left( a^{(1)}_{i,-n} -
                      q_1^{-n} a^{(1)}_{\overline{i+1},-n} \right) \right) \right] \times \\
                \times \exp \left[ - \sum_{n \geq 1}
                  \frac{z^{-n}}{n}q_2^{-n} ( a^{(2)}_{i,n} - q_3^{-n}
                  a^{(2)}_{\overline{i+1},n} )\right].
\end{multline}
The operators~\eqref{eq:74} should reproduce the web
$W$-algebras~\cite{gluing} associated with brane diagram of the form
shown in Fig.~\ref{fig:1}.

\subsection{Networks and partition functions}
\label{sec:netw-part-funct}
The network partition functions are built from the screening charges
in the same way as in the
$U_{q_1,q_2,q_3}(\widehat{\widehat{\mathfrak{gl}}}_1)$ case: one draws
the solid ``warp threads'' and then connects them with arbitrary
number of blue and red ``weft threads''. Let us give a simple example
of the construction:
\begin{multline}
  \label{eq:66}
    \includegraphics[valign=c]{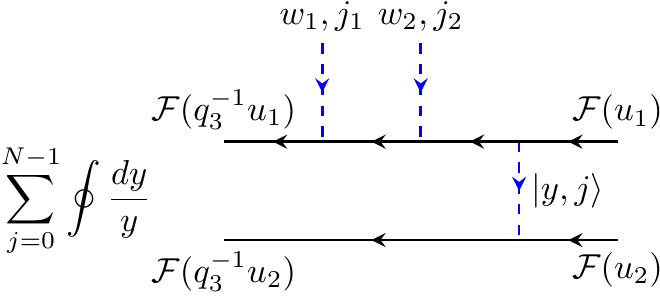} \quad =
    \sum_{j=0}^{N-1} \oint \frac{dy}{y} \begin{array}{c}
     \langle \varnothing |\\
    \otimes\\
    \langle \varnothing |
  \end{array}
    \Phi^{j_1}_{q_1}(w_1) \Phi^{j_2}_{q_1}(w_2) S_{q_1}^j(y)  \begin{array}{c}
    |\varnothing \rangle\\
    \otimes\\
    |\varnothing \rangle
  \end{array} \sim\\
  \sim (w_1 - w_2)^{\delta_{j_1,j_2}} \frac{\left( q_1^{\overline{j_2 - j_1}} q_2^{-1} \frac{w_2}{w_1} ; q_1^N
    \right)_{\infty}}{\left( q_1^{\overline{j_2 - j_1}}
      \frac{w_2}{w_1} ; q_1^N \right)_{\infty}} \sum_{j=0}^{N-1} \oint
  \frac{dy}{y}  y^{\log_{q_2} \frac{u_2}{u_1}} \prod_{a=1}^2 (w_a - q y )^{-\delta_{j_a,j}} \frac{\left( q_1^{\overline{j - j_a}} q \frac{y}{w_a} ; q_1^N
    \right)_{\infty}}{\left( q_1^{\overline{j - j_a}} q^{-1}
      \frac{y}{w_a} ; q_1^N \right)_{\infty}} .
\end{multline}
The integral~\eqref{eq:66} reproduces the partition function of the
\emph{orbifolded} $3d$ $\mathcal{N}=2^{*}$ $U(1)$ theory with two
hypermultiplets. The logic of this identification is similar
to~\cite{Z-higgsed}. One views the picture~\eqref{eq:66} as a brane
diagram of Type IIB string theory, so that the horizontal lines are
identified with NS5 branes and the vertical dashed lines are D3
branes. The finite D3 brane segment gives rise to a $U(1)$ gauge
theory, while the semi-infinite D3 branes represent the
hypermultiplets. The remaining directions of the D3 and NS5 branes are
then subject to the $\mathbb{Z}_N$ orbifolding. The FI, mass and
equivariant parameters of the $3d$ theory are identified with the
spectral parameters of the branes: $\log_{q_1} \frac{u_1}{u_2}$ is the
FI parameter, $w_{1,2}$ are exponentiated masses of the fundamental
hypermultiplets, $q_2$ is the real adjoint mass, and $q_1$ is the
equivariant parameter of the $3d$ $\Omega$-background.

A more general example is that of $U(n)$ theory with $m$ fundamental
hypermultiplets. To get the partition function we just add more dashed
lines to the diagram in~\eqref{eq:66}:
\begin{multline}
  \label{eq:68}
 \includegraphics[valign=c]{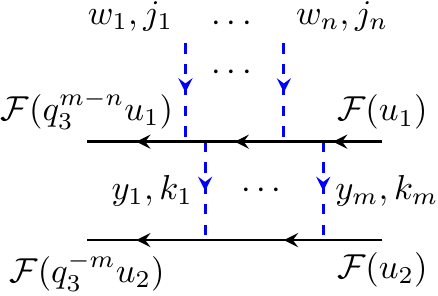} \quad \sim
  \prod_{a< b}(w_a - w_b)^{\delta_{j_a,j_b}} \frac{\left(
      q_1^{\overline{j_b - j_a}} q_2^{-1} \frac{w_b}{w_a} ; q_1^N
    \right)_{\infty}}{\left( q_1^{\overline{j_b - j_a}}
      \frac{w_b}{w_a} ; q_1^N \right)_{\infty}} \times\\
  \times \sum_{\{k_a \}=0}^{N-1} \oint d^ny \prod_{k=1}^N
  y_k^{\log_{q_2} \frac{u_2}{u_1}} \Delta_{\mathbb{Z}_N}^{q_1,
    q_2}(\vec{y}, \vec{k}) \prod_{a=1}^m \prod_{b=1}^n (w_a - q y_b
  )^{-\delta_{j_a,k_b}} \frac{\left( q_1^{\overline{k_b - j_a}} q
      \frac{y_b}{w_a} ; q_1^N \right)_{\infty}}{\left(
      q_1^{\overline{k_b - j_a}} q^{-1} \frac{y_k}{w_a} ; q_1^N
    \right)_{\infty}},
\end{multline}
where the ``orbifolded'' $(q_1,q_2)$-deformed Vandermonde determinant
is given by
\begin{equation}
  \label{eq:69}
  \Delta_{\mathbb{Z}_N}^{q_1,
    q_2}(\vec{y}, \vec{k}) = \prod_{i \neq j} (y_i - q_2
  y_j)^{\delta_{i,j}} \frac{\left( q_1^{\overline{k_i - k_j}}
      \frac{y_i}{y_j} ;q_1^N \right)_{\infty}}{\left(
      q_1^{\overline{k_i - k_j}} q_2 \frac{y_i}{y_j} ;q_1^N \right)_{\infty}}.
\end{equation}

Of course, one can add the blue and red dashed lines together to the
picture and obtain partition functions of two orbifolded $3d$ theories
living respectively on
\begin{equation}
  \label{eq:70}
  \mathbb{C}_{q_1}/\mathbb{Z}_N \times S^1 \subset
  \mathbb{C}^2_{q_1, q_3}/\mathbb{Z}_N \times S^1
\end{equation}
and
\begin{equation}
  \label{eq:71}
    \mathbb{C}_{q_3}/\mathbb{Z}_N \times S^1 \subset
  \mathbb{C}^2_{q_1, q_3}/\mathbb{Z}_N \times S^1
\end{equation}
and coupled through a $1d$ theory on $S^1$. We limit ourselves to a
simple example of two coupled $U(1)$ theories each with a single hypermultiplet:
\begin{multline}
  \label{eq:72}
  \includegraphics[valign=c]{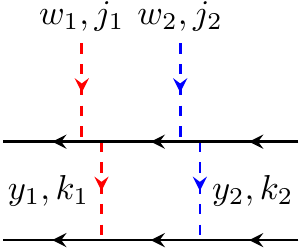}\qquad \sim (w_1 - q_1w_2)^{-\bar{\delta}_{j_1+1,j_2}}   \sum_{j=0}^{N-1} \oint
  d^2y (y_1 y_2)^{\log_{q_2} \frac{u_2}{u_1}} (y_1 -
  q_1y_2)^{-\bar{\delta}_{j_1+1,j_2}}\times\\
  \times (y_1 -
  q_3^{-1} y_2)^{-\bar{\delta}_{j_1+1,j_2}} (y_1 - d w_2
  )^{-\bar{\delta}_{k_1+1,j_2}} (y_2 - d w_1
  )^{-\bar{\delta}_{k_2+1,j_1}}\times\\
  \times (w_1 - q y_1
  )^{-\delta_{j_1,k_1}} \frac{\left( q_3^{\overline{j_1 - k_1}} q \frac{y_1}{w_1} ; q_3^N
    \right)_{\infty}}{\left( q_3^{\overline{j_1 - k_1}} q^{-1}
      \frac{y_1}{w_1} ; q_3^N \right)_{\infty}}  (w_2 - q y_2 )^{-\delta_{j_2,k_2}} \frac{\left( q_1^{\overline{k_2 - j_2}} q \frac{y_2}{w_2} ; q_1^N
    \right)_{\infty}}{\left( q_1^{\overline{k_2 - j_2}} q^{-1}
      \frac{y_2}{w_2} ; q_1^N \right)_{\infty}} .
\end{multline}

\section{Conclusions}
\label{sec:conclusions}
In this short note we have presented the natural generalization of the
Higgsed network formalism to ``non-abelian'' quantum toroidal algebras
$U_{q_1,q_2,q_3}(\widehat{\widehat{\mathfrak{gl}}}_N)$. We have found
the intertwiners (``Higgsed vertices''), computed their correlators,
and built screening operators from them. The correlators of
intertwiners are related to partition functions of $3d$ gauge theories
on orbifolded spacetimes.

There are several further questions about the models we have
considered. The partition functions we have obtained should satisfy a
version of $(q,t)$-KZ equation~\cite{Awata:2017lqa}, and they should
also be related to the eignefunctions of a spin generalization of the
Ruijsenaars-Schneider integrable system. It would be interesting to
clarify the interplay between these two systems.

An even more interesting, but also apparently more difficult task, is
to obtain general bosonization formulas for toroidal algebras. The
bosonization currently known for
$U_{q_1,q_2,q_3}(\widehat{\widehat{\mathfrak{gl}}}_N)$ is the Fock
representation~\cite{Saito-1,Saito-2}, which is written in terms of
$N$ free bosons and therefore has fixed central charge $C_1 = q$. It
is thus the toroidal analogue of the Frenkel-Kac
construction~\cite{Frenkel-Kac} for affine Lie algebras. This
bosonization also follows naturally from the geometric action of
$U_{q_1,q_2,q_3}(\widehat{\widehat{\mathfrak{gl}}}_N)$ on the moduli
space of instantons~\cite{Nagao}. A general bosonization formula for
$U_{q_1,q_2,q_3}(\widehat{\widehat{\mathfrak{gl}}}_N)$, if it exists,
should involve $N^2$ free bosons and produce arbitrary central
charges. This would be an analogue of the Wakimoto
bosonization~\cite{Wakimoto} for affine Lie algebras.

As we briefly explain in Appendix~\ref{sec:branching-into-}, quantum
toroidal algebras
$U_{q_1,q_2,q_3}(\widehat{\widehat{\mathfrak{gl}}}_N)$ are related to
toric Calabi-Yau three-folds of the form shown in
Fig.~\ref{fig:1}. More generally, one can consider an arbitrary toric
strip geometry, i.e.\ a manifold whose toric diagram has $N$
semi-infinite horizontal lines attached from the right and $M$ lines
attached from the left with no loops. This geometry corresponds to a
quantum toroidal \emph{superalgebra}
$U_{q_1,q_2,q_3}(\widehat{\widehat{\mathfrak{gl}}}_{N|M})$ recently
introduced in~\cite{BM}. It should be straightforward to generalize
our results to this case. An interesting conjecture is that \emph{any}
toric CY threefold corresponds to a quantum algebra, which is a
``two-dimensional'' generalization of quantum toroidal algebras. The
gluing approach recently popularized in~\cite{gluing} could probably
help to define these general algebras.

\paragraph{Acknowledgements.}
\label{sec:acknowledgements}
The author is partly supported by the RSF grant 18-71-10073.

\appendix

\section{The algebra
  $U_{q_1,q_2,q_3}(\widehat{\widehat{\mathfrak{gl}}}_N)$}
\label{sec:algebra}
In this Appendix we collect the definitions related to the algebra
$U_{q_1,q_2,q_3}(\widehat{\widehat{\mathfrak{gl}}}_N)$. We mainly
follow~\cite{FJMM-branching-rules}. To simplify the presentation we
limit ourselves to the cases when $N \geq 3$, since the $N=2$ case
requires special treatment. We use the standard notations for the
quantum parameters of the algebra\footnote{In~\cite{Z-higgsed} we have
  used a less symmetric notation
  $U_{q,t}(\widehat{\widehat{\mathfrak{gl}}}_1)$ for $U_{q_1, q_2,
    q_3}(\widehat{\widehat{\mathfrak{gl}}}_1)$, so that
  $q_{\cite{Z-higgsed}} = (q_1)_{\mathrm{here}}$ and
  $t_{\cite{Z-higgsed}} = (q_2)_{\mathrm{here}}^{-1}$.}
\begin{equation}
  \label{eq:1}
  q_1 = \frac{d}{q}, \qquad q_2 = q^2,\qquad q_3 = \frac{1}{q d}. 
\end{equation}
The algebra $U_{q_1,q_2,q_3}(\widehat{\widehat{\mathfrak{gl}}}_N)$ is
generated by modes of the currents:
\begin{gather}
  \label{eq:4}
  E_i(z) = \sum_{n \in \mathbb{Z}} E_{i,n} z^{-n}, \qquad F_i(z) =
  \sum_{n \in \mathbb{Z}} F_{i,n} z^{-n}, \\
  K^{\pm}_i(z) = K_i^{\pm 1} \pm \sum_{n \geq 1} K_{i,\pm n}
    z^{\mp n}  = K_i^{\pm 1}
  \exp \left[\pm (q - q^{-1}) \sum_{n \geq 1} H_{i,\pm n}
    z^{-n}\right]
\end{gather}
with $i = 0,\ldots, N-1$ and the central element $C_1$ satisfying the
following commutation relations\footnote{We conform with the
  conventions
  of~\cite{FJMM-branching-rules},~\cite{Awata:2017lqa}. The relations
  given in~\cite{FJMM-gln-reps},~\cite{Tsymbaliuk} are obtained from
  the convention we use by the redefinition
  $(K_i^{\pm}(z))_{\cite{FJMM-gln-reps},~\cite{Tsymbaliuk}} = (K_i^{\pm}(C_1^{1/2}
  z))_{\mathrm{here}}$.}:
\begin{gather}
  \label{eq:5}
  K_i^{\pm}(z) K_j^{\pm}(w) = K_j^{\pm}(w) K_i^{\pm}(z),\\
  \frac{g_{i,j}(C_1^{-1}z,w)}{g_{i,j}(C_1 z,w)} K_i^{-}(z) K_j^{+}(w)
  = \frac{g_{i,j}(w,C_1^{-1}z)}{g_{i,j}(w,C_1 z)} K_j^{+}(w)
  K_i^{-}(z),\\
  d_{i,j} g_{i,j}(z,w) K_i^{+} (z) E_j(w) + g_{j,i}(w,z) E_j(w)
  K_i^{+} (z) = 0,\\
  d_{i,j} g_{i,j}(z,w) K_i^{-} (C_1 z) E_j(w) + g_{j,i}(w,z) E_j(w)
  K_i^{-} (C_1 z) = 0,\\
  d_{j,i} g_{j,i}(w,z) K_i^{+} (C_1 z) F_j(w) + g_{i,j}(z,w) F_j(w)
  K_i^{+} (C_1 z) = 0,\\
  d_{j,i} g_{j,i}(w,z) K_i^{-} (z) F_j(w) + g_{i,j}(z,w) F_j(w)
  K_i^{-} (z) = 0,\\
  [E_i(z), F_j(z)] = \frac{\delta_{i,j}}{q-q^{-1}} \left( \delta
    \left( C_1 \frac{w}{z} \right) K^{+}_i(z) - \delta
    \left( C_1 \frac{z}{w} \right) K^{-}_i(w) \right),\\
  d_{i,j} g_{i,j}(z,w) E_i(z) E_j(w) + g_{j,i}(w,z) E_j(w) E_i(z) =
  0,\\
  d_{j,i} g_{j,i}(w,z) F_i(z) F_j(w) + g_{i,j}(z,w) F_j(w) F_i(z) =
  0,\\
  [E_i(z_1),[E_i(z_2), E_{i\pm 1}(w)]_q]_{q^{-1}} +
  [E_i(z_2),[E_i(z_1), E_{i\pm
    1}(w)]_q]_{q^{-1}} = 0,\\
  [F_i(z_1),[F_i(z_2), F_{i\pm 1}(w)]_q]_{q^{-1}} +
  [F_i(z_2),[F_i(z_1), F_{i\pm 1}(w)]_q]_{q^{-1}} = 0,
\end{gather}
where
\begin{align}
  \label{eq:6}
  [A,B]_q & \stackrel{\mathrm{def}}{=} AB - q BA, \\
  g_{i,j}(z, w) = (z - d^{-m_{i,j}} q^{a_{i,j}}w)&=
  \begin{cases}
    z- q_2 w, &i = j \bmod N,\\
    z- q_1 w, &i = j-1 \bmod N,\\
    z- q_3 w, &i = j+1 \bmod N,\\
    z - w, & \text{otherwise,}
  \end{cases}\\
  d_{i,j} = d^{m_{i,j}} &=
  \begin{cases}
    d^{\mp 1}, & i = j \mp 1 \bmod N,\\
    1, & \text{otherwise.}
  \end{cases}
\end{align}
The second central element of the algebra is given by the product of
the zero modes $K_i$:
\begin{equation}
  \label{eq:9}
  C_2 = \prod_{i=0}^{N-1} K_i.
\end{equation}
Sometimes it will be more convenient for us to write the relations
directly for the modes of the generating currents:
\begin{gather}
  \label{eq:10}
  [K_i, K_j] = 0,\quad \quad \qquad [K_i, H_{j,n}] = 0,\\
    K_i E_{j,n} = q^{a_{i,j}} E_{j,n} K_i,
  \quad \qquad  K_i F_{j,n} = q^{-a_{i,j}} F_{j,n} K_i,\\
[H_{i,n}, H_{j,m}] = \frac{d^{-n m_{i,j}}(q^{n a_{i,j}} -
  q^{-n a_{i,j}})(C_1^n - C_1^{-n})}{n(q - q^{-1})^2}  \delta_{n+m,0},\label{eq:31}\\
[H_{i,n}, E_{j,m}] =  d^{-n m_{i,j}} C_1^{\frac{n-|n|}{2}} \frac{(q^{n a_{i,j}} -
  q^{-n a_{i,j}})}{n(q - q^{-1})} E_{j,n+m},\\
[H_{i,n}, F_{j,m}] =  -d^{-n m_{i,j}} C_1^{\frac{n+|n|}{2}} \frac{(q^{n a_{i,j}} -
  q^{-n a_{i,j}})}{n(q - q^{-1})} F_{j,n+m},\\
[E_{i,n}, F_{j,m}] = \delta_{i,j} (1-\delta_{n+m,0}) \frac{K_{i,n+m}
  C_1^{-m \theta_{n+m>0} -n \theta_{n+m<0}}}{q-q^{-1}} + \delta_{i,j}
\delta_{n+m,0} \frac{C_1^n K_i - C_1^{-n} K_i^{-1}}{q - q^{-1}},\\
d^{m_{i,j}} ( E_{i,n+1} E_{j,m} - q^{a_{i,j}} E_{j,m} E_{i,n+1}) - (
  q^{a_{i,j}} E_{i,n} E_{j,m+1} - E_{j,m+1} E_{i,n}) = 0,\\
d^{m_{i,j}} (F_{i,n+1} F_{j,m} - q^{-a_{i,j}}  F_{j,m} F_{i,n+1}) - (q^{-a_{i,j}}F_{i,n} F_{j,m+1} -
   F_{j,m+1} F_{i,n}) = 0,\\
  [E_{i,n},[E_{i,m}, E_{i\pm 1,k}]_q]_{q^{-1}} +
  [E_{i,m},[E_{i,n}, E_{i\pm
    1,k}]_q]_{q^{-1}} = 0,\\
  [F_{i,n},[F_{i,m}, F_{i\pm 1,k}]_q]_{q^{-1}} +
  [F_{i,m},[F_{i,n}, F_{i\pm 1,k}]_q]_{q^{-1}} = 0,
\end{gather}
where $\theta_{n > 0}$ gives $1$ if $n >0$ or $0$ if $n \leq
0$.

The algebra is doubly graded with the first grading $d_1$ counting the
number of the mode of a current:
\begin{equation}
  \label{eq:14}
  [d_1, E_{i,n}] = n E_{i,n}, \qquad [d_1, F_{i,n}] = n F_{i,n},,
  \qquad [d_1, H_{i,n}] = n H_{i,n}, \qquad [d_1, K_i] = [d_1, C_1] = 0.
\end{equation}
The second grading $d_2$ is given by
\begin{equation}
  \label{eq:15}
  [d_2, E_{i,n}] = \delta_{i,0} E_{i,n}, \qquad [d_2, F_{i,n}] = -\delta_{i,0} F_{i,n},
  \qquad [d_2, H_{i,n}] = 0, \qquad [d_2, K_i] = [d_2, C_1] = 0.
\end{equation}

There are two natural quantum affine subalgebras
$U_q(\widehat{\mathfrak{gl}}_N)$ inside $U_{q_1,q_2,
  q_3}(\widehat{\widehat{\mathfrak{gl}}}_N)$, the ``horizontal'' and
the ``vertical'' ones. The horizontal subalgebra
$U_q(\widehat{\mathfrak{gl}}_N)_{\mathrm{hor}}$ is generated by the
zero modes of the currents $E_{i,0}$, $F_{i,0}$, $K_i$. Notice that
$i=0,\ldots, N-1$ labels the roots of the affine root system
$\widehat{A}_N$. This gives the ``traceless'' subalgebra
$U_q(\widehat{\mathfrak{sl}}_N)_{\mathrm{hor}}\subset
U_q(\widehat{\mathfrak{gl}}_N)_{\mathrm{hor}}$. An additional
horizontal Heisenberg subalgebra is generated by the sequential
commutators of non-zero modes, e.g.\ $E_{0,1}$ and $E_{i,-1}$ or
$E_{0,-1}$ and $E_{i,1}$. The horizontal subalgebra has vanishing
$d_1$ grading, while $d_2$ counts the modes of the quantum affine
currents.

The vertical quantum affine subalgebra
$U_q(\widehat{\mathfrak{gl}}_N)_{\mathrm{vert}}$ is obtained by
forgetting about the currents $E_0(z)$, $F_0(z)$. The remaining
currents $E_i(z)$, $F_i(z)$, $K^{\pm}_i(z)$ with $i=1,\ldots,N-1$ are
Drinfeld currents for the ``traceless part''
$U_q(\widehat{\mathfrak{sl}}_N)_{\mathrm{vert}}\subset
U_q(\widehat{\mathfrak{gl}}_N)_{\mathrm{vert}}$ of the vertical
subalgebra. Adding $K^{\pm}_0(z)$ gives an extra Heisenberg
subalgebra. The role of the gradings $d_1$ and $d_2$ is reversed in
the vertical subalgebra compared to the horizontal one.

\subsection{Coalgebra structure}
\label{sec:coproduct-structure}
The algebra $U_{q_1,q_2,q_3}(\widehat{\widehat{\mathfrak{gl}}}_N)$ can
be given a coalgebra structure with the coproduct structure given by
\begin{align}
  \label{eq:16}
  \Delta(E_i(z)) &= E_i(z) \otimes 1 + K^{-}_i(C_1^{(1)} z) \otimes E_i(C_1^{(1)} z),\\
  \Delta(F_i(z)) &= F_i(C_1^{(2)} z) \otimes K^{+}_i(C_1^{(2)} z) + 1 \otimes F_i(z),\\
  \Delta(K^{+}_i(z)) &= K^{+}_i(z) \otimes K^{+}_i((C_1^{(1)})^{-1}z),\label{eq:18}\\
  \Delta(K^{-}_i(z)) &= K^{-}_i((C_1^{(2)})^{-1} z) \otimes K^{-}_i(z),\label{eq:19}\\
  \Delta(C_1) &= C_1 \otimes C_1,\label{eq:17}
\end{align}
where $C_1^{(1)} = C_1 \otimes 1$, $C_1^{(2)} = 1 \otimes C_1 $. As in
the $U_{q_1,q_2,q_3}(\widehat{\widehat{\mathfrak{gl}}}_1)$ case, there
is an infinite number of equivalent but not identical coalgebra
structures, parametrized by splittings of the algebra into pairs of
Borel subalgebras. The coproducts corresponding to different coalgebra
structures are related by Drinfeld twists. For definiteness we call
the coproduct~\eqref{eq:16}--\eqref{eq:17} vertical, since it acts
(almost) diagonally on the vertical currents
$K_i^{\pm}(z)$~\eqref{eq:18},~\eqref{eq:19}.

\subsection{Symmetries of the algebra}
\label{sec:autom-algebra}
The algebra $U_{q_1,q_2, q_3}(\widehat{\widehat{\mathfrak{gl}}}_N)$
has a large group of symmetries including:
\begin{enumerate}
\item $SL(2,\mathbb{Z})$ automorphisms. The generator $S$ of this
  group is also known as the Miki's automorphism, which, in
  particular, transforms the horizontal quantum affine subalgebra into
  the vertical one. The central charges $C_1$ and $C_2$ and the
  gradings $d_1$ and $d_2$ are also exchanged by the automorphism
  $S$. The $SL(2,\mathbb{Z})$ duality group is shared by all known
  quantum toroidal algebras including the simplest one, $U_{q_1,q_2,
    q_3}(\widehat{\widehat{\mathfrak{gl}}}_1)$, because it originates
  from the mapping class group of the torus.
  
\item Reflection symmetry. This transformation, which we call $\rho$,
  exchanges the deformation parameters $q_1$ and $q_3$, and reflects
  the $\widehat{A}_N$ indices of the generating currents:
  \begin{gather}
    \rho: U_{q_1,q_2, q_3}(\widehat{\widehat{\mathfrak{gl}}}_N) \to U_{q_3,q_2, q_1}(\widehat{\widehat{\mathfrak{gl}}}_N),\notag\\
    \rho(E_i(z)) = E_{(-i) \bmod N}(z), \qquad \rho(F_i(z)) = F_{(-i)
      \bmod N}(z), \qquad \rho(H_i(z)) = H_{(-i) \bmod N}(z).    \label{eq:7}
\end{gather}
Evidently $\rho^2 = 1$, so the reflection symmetry generates
$\mathfrak{S}_2$. This symmetry is smaller, than the corresponding
symmetry of the $U_{q_1,q_2,
  q_3}(\widehat{\widehat{\mathfrak{gl}}}_1)$, in which case any the
permutation of $(q_1, q_2, q_3)$ was allowed (besides, no relabelling
of the generators was needed).

\item The symmetries listed so far were already present in the
  $U_{q_1,q_2,q_3}(\widehat{\widehat{\mathfrak{gl}}}_1)$ algebra. In
  the $\mathfrak{gl}_N$ case there is one more important automorphism
  $\sigma$, rotating the roots of $\widehat{A}_{N-1}$ root system:
  \begin{equation}
    \label{eq:8}
    \sigma(E_i(z)) = E_{i+1 \bmod N}(z), \qquad  \sigma(F_i(z))= F_{i+1 \bmod
      N}(z)\qquad   \sigma(K_i(z)) = K_{i+1 \bmod N}(z).
  \end{equation}
  Evidently, $\sigma^N = 1$. The central charges $C_1$ and $C_2$ are
  invariant under the symmetry, and so is the $d_1$ grading. The $d_2$
  grading is modified by the symmetry, but in a controlled way. Let
  $\delta_i$ be the grading associated with root $i$:
  \begin{equation}
    \label{eq:21}
    [\delta_i, E_{j,n}] = \delta_{i,j} E_{j,n},\qquad [\delta_i,
    F_{j,n}] = -\delta_{i,j} F_{j,n},\qquad  [\delta_i, H_{j,n}] = 0,
    \qquad  [\delta_i, K_j] = [\delta_i, C_1] = 0,
  \end{equation}
  so that $\delta_0 = d_2$. Then
  \begin{equation}
    \label{eq:20}
    \sigma(\delta_i) = \delta_{i+1 \bmod N}.
    \end{equation} 
\end{enumerate}

\subsection{The limit of $U_{q_1,q_2,
    q_3}(\widehat{\widehat{\mathfrak{gl}}}_N)$ to
  $W_{1+\infty}[\mathfrak{gl}_N]$ and the quantum torus}
\label{sec:algebra-as-quantum}
In the limit $q_2\to 1$ the algebra $U_{q_1,q_2,
  q_3}(\widehat{\widehat{\mathfrak{gl}}}_N)$ can be described very
explicitly as the central extension of the algebra of matrix-valued
functions on the quantum torus~\cite{FJMM-branching-rules}. The
quantum torus $T^2_{d^N}$ is a non-commutative space with
coordinates\footnote{We always assume that the coordinates are
  complexified.} $x$ and $p$ satisfying
\begin{equation}
  \label{eq:26}
  px = d^{-N} xp
\end{equation}
with $d = q_1 = q_3^{-1}$ playing the role of the parameter of the
noncommutativity. The algebra of matrix-valued functions is generated
by
\begin{equation}
  \label{eq:24}
  W^{(n,m)}_{i,j} = d^{-N\frac{nm}{2}} e_{i,j} x^n p^m, \qquad m,n \in
  \mathbb{Z}, \qquad i,j = 1,\ldots, N,
\end{equation}
where $e_{i,j}$ is a basis in the space of $N\times N$ matrices:
\begin{equation}
  \label{eq:23}
  (e_{i,j})_{k,l} = \delta_{i,k} \delta_{j,l}\qquad i,j,k,l =
  1,\ldots, N.
\end{equation}
After the central extension with two central charges $c_1$ and $c_2$,
the generators $W^{(n,m)}_{i,j}$ satisfy the following Lie algebra
relations:
\begin{equation}
  \label{eq:25}
  [W^{(n,m)}_{i,j}, W^{(r,s)}_{k,l}] = \delta_{j,k} d^{-N
    \frac{mr-ns}{2}} W^{(n+r,m+s)}_{i,l} - \delta_{i,l} d^{N
    \frac{mr-ns}{2}} W^{(n+r,m+s)}_{k,j} + \delta_{n+r,0}
  \delta_{m+s,0} \delta_{j,k} \delta_{i,l} (n c_1 + m c_2).
\end{equation}
Notice that the relations~\eqref{eq:25} are explicitly covariant under
the $SL(2,\mathbb{Z})$ transformations acting on the upper indices of
the generators $W^{(n,m)}_{i,j}$, while the central charges $c_1$ and
$c_2$ transform as a doublet. The algebra~\eqref{eq:25} is also called
$W_{1+\infty}[\mathfrak{gl}_N]$.

The generating currents of $U_{q_1,q_2,
  q_3}(\widehat{\widehat{\mathfrak{gl}}}_N)$ are given by the formulas
\begin{align}
  \label{eq:22}
  E_{i,n} &=
  \begin{cases}
    W_{i,i+1}^{(n,0)} d^{-in}, & i = 1,\ldots, N-1,\\
    W_{N,1}^{(n,1)} d^{-n \frac{N}{2}}, & i = 0,
  \end{cases}\\
  F_{i,n} &=
  \begin{cases}
    W_{i+1,i}^{(n,0)} d^{-in}, & i = 1,\ldots, N-1,\\
    W_{1,N}^{(n,-1)} d^{-n \frac{N}{2}}, & i = 0,
  \end{cases}\\
    H_{i,n} &=
  \begin{cases}
    (W_{i,i}^{(n,0)} -W_{i+1,i+1}^{(n,0)}) d^{-in}, & i = 1,\ldots, N-1,\\
    (d^{-n N} W_{N,N}^{(n,0)} - W_{1,1}^{(n,0)}) + c_2 \delta_{n,0}, &
    i = 0,
  \end{cases}\\
  c_{1,2} &= \lim_{q_2 \to 1} \log_{q_2} C_{1,2}. 
\end{align}

\subsection{Vector representations of
  $U_{q_1,q_2,q_3}(\widehat{\widehat{\mathfrak{gl}}}_N)$}
\label{sec:vect-repr-u}
Vector representation $\mathcal{V}_{q_1}$ is a representation of
$U_{q_1,q_2,q_3}(\widehat{\widehat{\mathfrak{gl}}}_N)$ on the space of
$\mathbb{C}^N$-valued functions of a single variable $x$. It has
trivial central charges, $C_1 = C_2 = 1$. We can take as basis states
of this representation tensor products of Dirac delta functions and
basis vectors $\mathbf{e}_j$ in $\mathbb{C}^N$. We write them
as\footnote{Our notation for the states of the vector representation
  is related to that of~\cite{FJMM-branching-rules} by the formula
  \begin{equation}
    \left([u]_j^{(k)}\right)_{\cite{FJMM-branching-rules}} =
    \left(|q_1^{j-k} u, (k-j-1) \bmod N +
      1\rangle \right)_{\mathrm{here}}, \qquad j \in \mathbb{Z},\notag  
  \end{equation}
  so that
  \begin{align}
    \label{eq:36}
    E_i(z)[u]_j^{(k)}&= \delta \left( q_1^{j+1} \frac{u}{z} \right)
      \bar{\delta}_{i+j+1,k} [u]_{j+1}^{(k)},\\
    F_i(z)[u]_j^{(k)}&= \delta \left( q_1^j \frac{u}{z} \right)
      \bar{\delta}_{i+j,k} [u]_{j-1}^{(k)},\\
    K^{\pm}_i(z)[u]_j^{(k)}&=\psi^{\bar{\delta}_{i+j,k}} \left( q_1^j
          \frac{u}{z} \right) \psi^{-\bar{\delta}_{i+j+1,k}} \left(
              q_1^j q_3^{-1} \frac{u}{z} \right) [u]_j^{(k)}.
  \end{align}
}
\begin{equation}
  \label{eq:30}
  |u, j\rangle = \delta \left( q_1^j \frac{x}{u} \right) \mathbf{e}_j,
  \qquad j = 1,\ldots, N.
\end{equation}
In Eq.~\eqref{eq:30} we view the spectral parameter $u$ as the
parameter of the state, not of the representations (the latter
convention was adopted in~\cite{FJMM-branching-rules},
\cite{FJMM-gln-reps}). These two views are in fact completely
equivalent. The algebra
$U_{q_1,q_2,q_3}(\widehat{\widehat{\mathfrak{gl}}}_N)$ acts on the
states~\eqref{eq:30} as matrices valued in difference
operators\footnote{Note that the action of the shift operator in $x$
  on a state $|u,j\rangle$ can be understood as an \emph{inverse}
  shift of the parameter $u$:
\begin{equation}
  q_1^{x \partial_x}
  |u,j\rangle = |q_1 ^{-1} u,j\rangle.\notag
\end{equation}} in $x$:
\begin{align}
  \label{eq:28}
  E_i(z) &=
  \begin{cases}
    e_{N,1} \delta \left( \frac{x}{z} \right) q_1^{-x \partial_x}, & i=0,\\
    e_{i,i+1} \delta \left( \frac{x}{z} \right) q_1^{-x \partial_x},
    &i=1,\ldots,N-1,
  \end{cases}\\
  F_i(z) &=
  \begin{cases}
    e_{1,N} \delta \left( q_1 \frac{x}{z} \right) q_1^{x \partial_x}, & i=0,\\
    e_{i+1,i} \delta \left( q_1 \frac{x}{z} \right) q_1^{x \partial_x},
    &i=1,\ldots,N-1,
  \end{cases}\\
  K^{\pm}_i(z) &= 
  \begin{cases}
    \psi^{e_{N,N}} \left( \frac{x}{z} \right) \psi^{-e_{1,1}}
    \left(q_3^{-1} \frac{x}{z} \right),
    &i=0,\\
    \psi^{e_{i,i}} \left( \frac{x}{z} \right) \psi^{-e_{i+1,i+1}}
    \left(q_3^{-1} \frac{x}{z} \right), &i=1,\ldots, N-1,
  \end{cases}\label{eq:2}
\end{align}
where
\begin{equation}
  \label{eq:29}
  \psi(x) = \sqrt{q_2} \frac{1 - q_2^{-1}x}{1- x} = q_2^{-\frac{1}{2}} \frac{1 - q_2x^{-1}}{1- x^{-1}},
\end{equation}
and $e_{i,j}$ are basis matrices defined in Eq.~\eqref{eq:23}. The
eigenvalues of the currents $K^{+}(z)$ and $K^{-}(z)$ are given by the
same rational function understood as a series expansion around $z=0$
and $z = \infty$ respectively. Let us also write the zero modes $K_i$
explicitly:
\begin{equation}
  \label{eq:13}
  K_i =
  \begin{cases}
    q_2^{\frac{e_{N,N} - e_{1,1}}{2}},
    &i=0,\\
    q_2^{\frac{e_{i,i} - e_{i+1,i+1}}{2}}, &i=1,\ldots, N-1,
  \end{cases}
\end{equation}
so that indeed $C_2 = \prod_{i=0}^{N-1} K_i = 1$.

Due to the symmetry $\rho$ of the algebra (Eq.~\eqref{eq:7}) there
exists a representation $\mathcal{V}_{q_3}$ defined similarly to
Eq.~\eqref{eq:28}--\eqref{eq:2} with $q_1 \leftrightarrow q_3$ and the
labels of the generators reversed.

In~\cite{FJMM-branching-rules} a set of $N$ different vector
representations $\mathcal{V}_{q_1}^{(k)}$ related to each other by the
action of $\sigma$ has been defined. Due to $\sigma$ being the
symmetry of the algebra, all the representations in the family are
isomorphic. Here we make do with just one representation
$\mathcal{V}_{q_1}$. We do so by noticing that $\sigma$ can be
represented on the space of vector-valued functions as a cyclic
permutation matrix $s$:
\begin{equation}
  \label{eq:11}
  \sigma (g) |u,j\rangle = s g s^{-1} |u,j\rangle,
\end{equation}
where
\begin{equation}
  \label{eq:12}
  s = \sum_{i=1}^{N-1} e_{i+1,i} + e_{1,N}.
\end{equation}
In this way all $N$ representations $\mathcal{V}_{q_1}^{(k)}$ are
obtained from $\mathcal{V}_{q_1} = \mathcal{V}_{q_1}^{(0)}$ by the
shift of the indices.

\subsection{Horizontal Fock representation of
  $U_{q_1,q_2,q_3}(\widehat{\widehat{\mathfrak{gl}}}_N)$}
\label{sec:fock-representation-}
In the horizontal Fock representation representation~\cite{Saito-1,
  Saito-2} $\mathcal{F}^{(1,0),k}_{q_1,q_3}(u)$ all currents of the
$U_{q_1,q_2,q_3}(\widehat{\widehat{\mathfrak{gl}}}_N)$ algebra are
expressed through vertex operators built from the generators $H_{i,n}$
constituting $N$ copies of the Heisenberg algebra (see
Eq.~\eqref{eq:31}). The central charges of
$\mathcal{F}^{(1,0),k}_{q_1,q_3}(u)$ are $C_1 = q = \sqrt{q_2}$ and
$C_2 = 1$. The representation $\mathcal{F}^{(1,0),k}_{q_1,q_3}(u)$ is
characterized by a spectral parameter $u \in \mathbb{C}^{\times}$ and
the color index $k = 0, \ldots, N-1$.

We define the fundamental vertex operators
\begin{equation}
  \label{eq:27}
  V^{\pm}_i(z) = \exp \left[\mp \sum_{n \geq 1} \frac{q-q^{-1}}{q^n -
      q^{-n}} H_{i,\pm n} z^{\mp n} \right].
\end{equation}
Notice that $V^{\pm}(z)$ contain only positive or negative modes
respectively and are therefore automatically normal ordered. The
product of $V^{-}(z) V^{+}(w)$ is normal ordered too, but $V^{+}(z)
  V^{-}(w)$ is not, and for it we have the following identity:
\begin{equation}
V_{i}^{(+)} (z) V_{j}^{(-)}(w) = s_{ij}(z,w) V_{j}^{(-)}(w) V_{i}^{(+)} (z) ,  \qquad |z| > |w|,
\end{equation}
where 
\begin{equation}
s_{ij}(z,w) = 
\frac{\left( 1 - \frac{q w}{z} \right)^{\bar{\delta}_{i,j}} \left( 1 - \frac{w}{q z} \right)^{\bar{\delta}_{i,j}} }
{\left( 1 - \frac{d w}{z} \right)^{\bar{\delta}_{i,j-1}}\left( 1 - \frac{w}{ d z} \right)^{\bar{\delta}_{i,j+1}}}.
\end{equation}
where $\bar{\delta}_{i,j}$ gives one if $i=j \bmod N$ and zero
otherwise. Using $V^{\pm}(z)$ we define the vertex operator
representation of
$U_{q_1,q_2,q_3}(\widehat{\widehat{\mathfrak{gl}}}_N)$:
\begin{align}
  E_i(z) &= \eta_i(z)~e^{\bar{\alpha}_i} z^{H_{i,0}+1} e^{\bar{\delta}_{i,k}P},  \label{hE}\\
F_i(z) &= \xi_i(z)~e^{- {\bar{\alpha}_i}} z^{- H_{i,0}+1}  e^{-\bar{\delta}_{i,k}P},  \label{hF} \\
K_i^{\pm}(q^{\frac{1}{2}} z) &= \varphi_i^{\pm}(z)~q^{\pm H_{i,0}}. \label{hK}  
\end{align}
where $P$ is the momentum operator giving $\ln u$ on the whole space
$\mathcal{F}^{(1,0),k}_{q_1,q_3}(u)$ (this definition seems
superfluous when we are dealing with a single Fock representation, but
will be convenient when we will study tensor products of Fock
representations). The vertex operators $\eta_i(z)$, $\xi_i(z)$ and
$\phi^{\pm}_i(z)$ are given by:
\begin{align}
   \eta_i(z) &= V_{i}^{-} ( q^{-\frac{1}{2}} z ) V_{i}^{+} (q^{\frac{1}{2}} z), \label{eq:41}\\
    \xi_i(z) &= (V_{i}^{-} ( q^{\frac{1}{2}} z ))^{-1} (V_{i}^{+} (q^{-\frac{1}{2}} z))^{-1}, \\
      \varphi_i^{\pm}(z) &= V_{i}^{\pm}(q^{\pm1} z) (V_{i}^{\pm}
      (q^{\mp 1} z))^{-1}.\label{eq:42}
\end{align}
and $\bar{\alpha}_i$, $H_{i,0}$ ($i=0,\ldots, N-1$) are the zero modes
satisfying the relations\footnote{Our relations for the zero modes
  differ slightly from the relations found
  in~\cite{Saito-1,Saito-2,Awata:2017lqa}. There \emph{three} types of
  zero modes $\bar{\alpha}_i$, $H_{i,0}$, $\partial_{\bar{\alpha}_i}$
  satisfying
\begin{align}
 e^{\bar{\alpha}_i} e^{\bar{\alpha}_j}  &= (-1)^{a_{i,j}}  e^{\bar{\alpha}_j} e^{\bar{\alpha}_i},  \\
\cite{Saito-1,Saito-2,Awata:2017lqa}:\qquad z^{\partial_{\bar{\alpha}_i}} e^{\bar{\alpha}_j} &= z^{a_{i,j}}e^{\bar{\alpha}_j} z^{\partial_{\bar{\alpha}_i}}, \\
z^{H_{i,0}} e^{\bar{\alpha}_j} &= z^{a_{i,j}} d^{-\frac{1}{2} m_{i,j}} e^{\bar{\alpha}_j} z^{H_{i,0}} \label{eq:43}
\end{align}
have been used. Notice that the power notation $z^{H_{i,0}}$ in this
case is somewhat misleading, since Eq.~\eqref{eq:43} is inconsistent
with $z_1^{H_{i,0}}z_2^{H_{i,0}} = (z_1 z_2)^{H_{i,0}}$.}
\begin{align}
  e^{\bar{\alpha}_i} e^{\bar{\alpha}_j}  &= (-d)^{-m_{i,j}}  e^{\bar{\alpha}_j} e^{\bar{\alpha}_i}, \label{cocycle} \\
z^{H_{i,0}} e^{\bar{\alpha}_j} &= z^{a_{i,j}} e^{\bar{\alpha}_j}
z^{H_{i,0}}, \\
[H_{i,0}, H_{j,0}] &= 0.\label{momentum}
\end{align}
The relation~\eqref{cocycle} can be understood as a
$\mathbb{C}^{\times}$ central extension of the group algebra of the
$\widehat{A}_{N-1}$ root lattice\footnote{This is an obvious
  generalization the $\mathbb{Z}_2$ central extension, which is used
  in conventional vertex operator algebras.}. Notice that
$\sum_{i=0}^{N-1} \bar{\alpha}_i$ and $\sum_{i=0}^{N-1} H_{i,0}$ are
central elements of the zero mode
algebra~\eqref{cocycle}--\eqref{momentum}. Since for
$\mathcal{F}^{(1,0),k}_{q_1,q_3}(u)$ the central charge $C_2 = 1$, we
have to fix
\begin{equation}
  \label{eq:44}
     \sum_{i=0}^{N-1} H_{i,0} = 0.
\end{equation}
We also set
\begin{equation}
  \label{eq:46}
  \sum_{i=0}^{N-1}
\bar{\alpha}_i = 0.
\end{equation}
Due to Eqs.~\eqref{eq:44},~\eqref{eq:46} we can understand $H_{i,0}$
as measuring discrete ``momenta'' taking values in the weight lattice
of $A_{N-1}$ and $e^{\bar{\alpha}_i}$ as the generators shifting these
momenta by simple roots of $A_{N-1}$.

Thus, the states in the Fock representation are generated by $N$ free
bosons $H_{i,n}$ for $n \neq 0$, and in addition by the zero modes
$e^{\bar{\alpha}_i}$. A general state is labelled by an $N$-tuple of
Young diagrams $\vec{Y}$ and a weight $\bar{\mu}$ of the algebra
$\mathfrak{sl}_N$. We denote this state by $|u, e^{\bar{\mu}},
\vec{Y}\rangle$. The vacuum state of the representation $\mathcal{F}^{(1,0),k}_{q_1,q_3}(u)$ is
\begin{equation}
  \label{eq:53}
  |u , e^{\bar{\omega}_k}, \vec{\varnothing} \rangle
\end{equation}
where $\bar{\omega}_k$ is the $k$-th fundamental weight of
$A_{N-1}$.

\paragraph{Independent Heisenberg generators.}
\label{sec:indep-heis-gener}
We find it convenient to introduce another basis $a_{i,n}$ in $N$
copies of Heisenberg algebra spanned by $H_{i,n}$ ($i=
0,\ldots,N-1$). We write
\begin{equation}
  \label{eq:40}
  H_{i,n} = \frac{(q^n -
    q^{-n})}{n (q - q^{-1})} q^{- \frac{|n|}{2}}  \left( a_{i,n} - q^{|n|} d^{n} a_{\overline{i+1}, n}  \right),
\end{equation}
where $\overline{i} = i \bmod N$, so that bosons $a_{i,n}$ are
completely independent:
\begin{equation}
  \label{eq:3}
  [a_{i,n}, a_{j,m}] = n \delta_{n+m,0} \delta_{i,j} .
\end{equation}

We can also simplify the relations~\eqref{cocycle},~\eqref{momentum}
by introducing zero modes $Q_i$, $P_i$, satisfying the canonical
commutation relations
\begin{equation}
  \label{eq:47}
  [P_i, Q_j]= \delta_{i,j}, \qquad [Q_i, Q_j] = [P_i, P_j] = 0, \qquad
  i,j = 0,\ldots, N-1.
\end{equation}
The zero modes $e^{Q_i}$, $H_{i,0}$ are expressed through $Q_i$, $P_i$
as follows:
\begin{equation}
   \label{eq:48}
   e^{\bar{\alpha}_i} =  e^{Q_i - Q_{\overline{i+1}}}
       (-d)^{P_{\overline{i+1}}},\qquad H_{i,0} = P_i
         - P_{\overline{i+1}}.
\end{equation}

In terms of $a_{i,n}$, $Q_i$, $P_i$ we have the following expression
for the vertex operators~\eqref{eq:41}--\eqref{eq:42} (similar
formulas appeared in~\cite{tor-spectral}):
\begin{align}
  E_i(z) &= e^{\delta_{i,k}P} e^{Q_i - Q_{\overline{i+1}}} z^{P_i -
        P_{\overline{i+1}}+1} (-d)^{P_{\overline{i+1}}} \exp \left[ \sum_{n \geq 1}
                  \frac{z^n}{n}q^{-n} ( a_{i,-n} - q_1^{-n} a_{\overline{i+1},-n} )\right]\times \notag \\
            &\phantom{=} \times \exp \left[ - \sum_{n \geq 1}
              \frac{z^{-n}}{n}q^{-n} ( a_{i,n} - q_3^{-n} a_{\overline{i+1},n} )\right],  \label{eq:33}\\
            F_i(z) &= e^{-\delta_{i,k}P} e^{-Q_i + Q_{\overline{i+1}}} z^{-P_i + P_{\overline{i+1}}+1} (-d)^{-P_{\overline{i+1}}} \exp \left[ - \sum_{n \geq 1} \frac{z^n}{n}(a_{i,-n} - q_1^{-n} a_{\overline{i+1},-n}
                )\right]\times \notag \\
              &\phantom{=} \times \exp \left[\sum_{n \geq 1} \frac{z^{-n}}{n}(a_{i,n} - q_3^{-n} a_{\overline{i+1},n}
                    )\right],\\
                  K^{+}_i(z) &= q^{P_i - P_{\overline{i+1}}} \exp \left[ \sum_{n \geq 1} \frac{z^{-n}}{n} (q^n - q^{-n})q^{-\frac{n}{2}} \left( a_{i,n}
                        - q_3^{-n} a_{\overline{i+1},n} \right)\right],\\
                        K^{-}_i(z) &= q^{-P_i + P_{\overline{i+1}}} \exp \left[ - \sum_{n \geq 1} \frac{z^n}{n} (q^n - q^{-n})q^{-\frac{n}{2}} \left( a_{i,-n} - q_1^{-n} a_{\overline{i+1},-n} \right)\right].
\end{align}

Apparently, for $U_{q_1,q_2,q_3}(\widehat{\widehat{\mathfrak{gl}}}_N)$
only one type of Fock representations
$\mathcal{F}^{(1,0),k}_{q_1,q_3}(u)$ with central charge $C_1 =
\sqrt{q_2}$ is known. $\mathcal{F}^{(1,0),k}_{q_1,q_3}(u)$ is
invariant under the reflection symmetry $\rho$~\eqref{eq:7}. The
rotation symmetry $\sigma$ shifts the color index $k$ of the
representations by one modulo $N$. This situation is in contrast to
$U_{q_1,q_2,q_3}(\widehat{\widehat{\mathfrak{gl}}}_1)$, where there
where \emph{three} Fock representations
$\mathcal{F}^{(1,0),k}_{q_a,q_b}(u)$ with $a\neq b$ and central
charges $C_1 = \sqrt{q_c}$, where $c \neq a \neq b$. Of course, since
$q_2$ plays a distinguished role in the algebra
$U_{q_1,q_2,q_3}(\widehat{\widehat{\mathfrak{gl}}}_N)$, the fact that
the representation with central charge related to $q_2$ is also
distinguished in this case is not surprising. Still, we think that
this point deserves further investigation.

\subsection{Branching into $U_{q_1, q_2,
    q_3}(\widehat{\widehat{\mathfrak{gl}}}_1)$ algebras and toric
  geometry}
\label{sec:branching-into-}
\begin{figure}[h]
  \centering
  \includegraphics{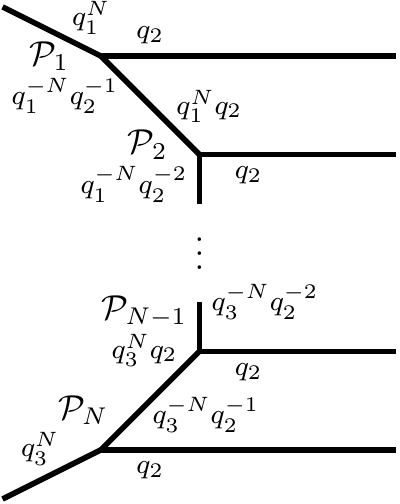}
  \caption{Toric diagram of the threefold $X_N \times \mathbb{C}$,
    where $X_N = \widetilde{\mathbb{C}^2/\mathbb{Z}_N}$. Each
    trivalent vertex corresponds to a $(\mathbb{C}^{\times})^3$ fixed
    point $\mathcal{P}_a$ with indices on the legs denoting the
    weights of the $(\mathbb{C}^{\times})^3$ action. These weights,
    $q^{(a)}_i$ enter as the parameters of the $U_{q^{(a)}_1,
      q^{(a)}_2, q^{(a)}_3}(\widehat{\widehat{\mathfrak{gl}}}_1)$
    subalgebra of $U_{q_1, q_2,
      q_3}(\widehat{\widehat{\mathfrak{gl}}}_N)$.}
  \label{fig:1}
\end{figure}

There is a geometric way of viewing the algebra $U_{q_1, q_2,
  q_3}(\widehat{\widehat{\mathfrak{gl}}}_N)$, which can be helpful for
understanding its representations and subalgebras. The algebra
$U_{q_1, q_2, q_3}(\widehat{\widehat{\mathfrak{gl}}}_N)$ is associated
with a toric Calabi-Yau three-fold of the form $X_N \times
\mathbb{C}$, where $X_N = \widetilde{\mathbb{C}^2/\mathbb{Z}_N}$ is
the blowup of the $A_N$ singularity. There is a $\mathbb{C}^3$ action
on the toric CY; denote its weights by $q_1$, $q_2$, $q_3$. We turn
off the $\mathbb{C}^{\times}$ part of $\mathbb{C}^3$ which scales the
canonical bundle, so the weights satisfy $q_1 q_2 q_3 = 1$. We choose
the weights so that $q_2$ scales the trivial $\mathbb{C}$ fiber, while
the choice of $q_1$ is easier to shown on the toric diagram
(Fig.~\ref{fig:1}).

Let us notice that for $N=1$ the threefold is simply $\mathbb{C}^3$
with weights $q_i$ rotating the coordinate planes $\mathbb{C}$. It is
therefore logical to assume that the neighbourhoods of the fixed
points $\mathcal{P}_a$ of the $(\mathbb{C}^{\times})^3$ action inside
$X_N \times \mathbb{C}$ might be related to $U_{q^{(a)}_1, q^{(a)}_2,
  q^{(a)}_3}(\widehat{\widehat{\mathfrak{gl}}}_1)$ subalgebras inside
$U_{q_1, q_2, q_3}(\widehat{\widehat{\mathfrak{gl}}}_N)$. Having
accepted this hypothesis we see that the deformation parameters
$q^{(a)}_i$ should be related to the weights of the fixed planes
joining at the fixed point $a$. They are fixed by the geometry of the
threefold (see Fig.~\ref{fig:1}):
\begin{equation}
  \label{eq:34}
  q_1^{(a)} = q_1^N q_2^{a-1}, \qquad q_2^{(a)} = q_2, \qquad
  q_3^{(a)} = q_1^{-N} q_2^{-a}, \qquad a = 1,\ldots, N.
\end{equation}
These turn out to be the correct parameters of the $U_{q^{(a)}_1,
  q^{(a)}_2, q^{(a)}_3}(\widehat{\widehat{\mathfrak{gl}}}_1)$
subalgebras of $U_{q_1, q_2,
  q_3}(\widehat{\widehat{\mathfrak{gl}}}_N)$ described
in~\cite{FJMM-branching-rules}. These subalgebras are easier to
understand in the $q_2 \to 1$ limit. In this limit, described in
Appendix~\ref{sec:algebra-as-quantum}, the algebra $U_{q_1, q_2,
  q_3}(\widehat{\widehat{\mathfrak{gl}}}_N)$ turns into the central
extension of the algebra of matrix-valued functions on the quantum
torus $T^2_{d^N}$. The subalgebras $U_{q^{(a)}_1, q^{(a)}_2,
  q^{(a)}_3}(\widehat{\widehat{\mathfrak{gl}}}_1)$ is generated by
functions on $T^2_{d^N}$ valued in the \emph{diagonal} matrices:
\begin{equation}
  \label{eq:38}
  U_{q^{(a)}_1, q^{(a)}_2,
    q^{(a)}_3}(\widehat{\widehat{\mathfrak{gl}}}_1) =  \langle
  W^{(n,m)}_{a, a} \rangle, \qquad n,m \in \mathbb{Z}.
\end{equation}
Notice also that in this limit $q^{(a)}_i$ are independent
of $a$:
\begin{equation}
  \label{eq:37}
  q_1^{(a)} = q_1^N = d^N, \qquad q_2^{(a)} = 1, \qquad
  q_3^{(a)} = q_1^{-N} = d^{-N}, \qquad \text{for } q_2 \to 1. 
\end{equation}

One can also deduce some larger subalgebras of $U_{q_1, q_2,
  q_3}(\widehat{\widehat{\mathfrak{gl}}}_N)$ using the geometric
picture. Indeed, let us cut the lowest vertex from the diagram in
Fig.~\ref{fig:1}. We are left with the CY threefold $X_{N-1} \times
\mathbb{C}$, which inherits the $(\mathbb{C}^{\times})^3$ action from $X_N \times
\mathbb{C}$. The corresponding subalgebra of $U_{q_1, q_2,
  q_3}(\widehat{\widehat{\mathfrak{gl}}}_N)$ is $U_{q'_1, q'_2,
  q'_3}(\widehat{\widehat{\mathfrak{gl}}}_{N-1})$ with
\begin{equation}
  \label{eq:39}
  q_1' = q_1^{1 + \frac{1}{N-1}}, \qquad q_2' = q_2, \qquad q_3' =
  q_3 q_1^{-\frac{1}{N-1}}.
\end{equation}
This subalgebra in the $q_2 \to 1$ limit corresponds to $(N-1)\times
(N-1)$ matrix-valued functions on the quantum torus with the
$(N-1)\times (N-1)$ block embedded into $N \times N$ matrices in the
upper left corner. More generally by cutting the toric diagram from
Fig.~\ref{fig:1} one can get subalgebras corresponding to $k \times k$
blocks inside an $N \times N$ matrix sitting on the diagonal. The
parameters are easily deduces from the picture.

Thus the structure of the toric skeleton of the CY threefold is
related to certain natural subalgebras of $U_{q_1, q_2,
  q_3}(\widehat{\widehat{\mathfrak{gl}}}_N)$. The correspondence we
have described originates from the geometric representation theory. In
particular, the $U_{q_1, q_2,
  q_3}(\widehat{\widehat{\mathfrak{gl}}}_N)$ algebra acts by
correspondences on the equivariant $K$-theory of the moduli space of
instantons on $X_N$. The correspondences can be seen as adding or
removing instantons from the theory. In the equivariant setup the
instantons are effectively of zero size and concentrate near fixed
points of the equivariant action. Hence, there are subsets of
correspondences adding or removing instantons only near \emph{some}
fixed points. These subsets turns out to be a subalgebras.

A more physical argument for cutting the toric diagram might be to
notice that the action on the $K$-theory is independent of the area of
the two-cycles inside $X_N$, so one can scale the area of some cycles
to become infinitely large, in which case the fixed points separated
by these two-cycles stop talking to each other.

One can view the reduction of $U_{q_1, q_2,
  q_3}(\widehat{\widehat{\mathfrak{gl}}}_N)$ to $ \bigoplus_{a=1}^N
U_{q^{(a)}_1, q^{(a)}_2,
  q^{(a)}_3}(\widehat{\widehat{\mathfrak{gl}}}_1)$ in an opposite way:
the larger algebra can be ``glued'' from the sum of smaller ones by
adding the ``off-diagonal'' generators, which transform bimodules
under $U_{q^{(a)}_1, q^{(a)}_2,
  q^{(a)}_3}(\widehat{\widehat{\mathfrak{gl}}}_1)$ and $U_{q^{(b)}_1,
  q^{(b)}_2, q^{(b)}_3}(\widehat{\widehat{\mathfrak{gl}}}_1)$ for some
$a$ and $b$. The cohomological limit of such a gluing procedure has
been the subject of many recent works~\cite{gluing}.

\end{document}